\documentclass[twocolumn,superscriptaddress]{revtex4}
\usepackage{graphicx}
\usepackage{epsfig}
\usepackage{amsmath}
\usepackage{amsfonts,amsbsy}
\usepackage{amssymb}
\usepackage{hyperref}
\usepackage{bm}
\usepackage{color}
\usepackage{enumerate}
\usepackage{verbatim}
\usepackage{subfloat}
\usepackage{subfigure}
\def\be{\begin{equation}}
\def\ee{\end{equation}}
\def\bea{\begin{eqnarray}}
\def\eea{\end{eqnarray}}

\begin{document}

\title{Exploring percolation phase transition in the three-dimensional Ising model with machine learning }

\author{Ranran Guo}
\affiliation{Key Laboratory of Quark and Lepton Physics (MOE) and Institute of Particle Physics, \\
Central China Normal University, Wuhan 430079, China}
\author{Xiaobing Li}
\affiliation{Key Laboratory of Quark and Lepton Physics (MOE) and Institute of Particle Physics, \\
Central China Normal University, Wuhan 430079, China}
\author{Rui Wang}
\affiliation{Key Laboratory of Quark and Lepton Physics (MOE) and Institute of Particle Physics, \\
Central China Normal University, Wuhan 430079, China}
\author{Shiyang Chen}
\affiliation{Department of Physics, Swansea University, SA2 8PP, Swansea, United Kingdom}
\author{Yuanfang Wu}
\affiliation{Key Laboratory of Quark and Lepton Physics (MOE) and Institute of Particle Physics, \\
Central China Normal University, Wuhan 430079, China}
\author{Zhiming Li}
\email{lizm@mail.ccnu.edu.cn}
\affiliation{Key Laboratory of Quark and Lepton Physics (MOE) and Institute of Particle Physics, \\
Central China Normal University, Wuhan 430079, China}
\affiliation{College of Physics and Electronic Engineering, Hanjiang Normal University, Shiyan 442000, China}

\begin{abstract}
The percolation study offers valuable insights into the characteristics of phase transition, shedding light on the underlying mechanisms that govern the formation of global connectivity within the system. We explore the percolation phase transition in the 3D cubic Ising model by employing two machine learning techniques. Our results demonstrate the capability of machine learning methods in distinguishing different phases during the percolation transition. Through the finite-size scaling analysis on the output of the neural networks, the percolation temperature and a correlation length exponent in the geometrical percolation transition are extracted and compared to those in the thermal magnetization phase transition within the 3D Ising model. These findings provide  a valuable way essential for enhancing our understanding of the property of the QCD critical point, which belongs to the same universality class as the 3D Ising model.
\end{abstract}

\maketitle
%%%%%%%%%%%%%%%%%%%%%%%%%%%%%%%%%%%%%%%%%%%%%%%%%%%%%%%%%%%%%%%%%%%%%%%%%%%%%%
\section{Introduction}\label{sec:introduction}
%%%%%%%%%%%%%%%%%%%%%%%%%%%%%%%%%%%%%%%%%%%%%%%%%%%%%%%%%%%%%%%%%%%%%%%%%%%%%%
Exploring the phase diagram of Quantum Chromodynamics (QCD) and searching for the critical point (CP) of the phase transition from hadrons to Quark-Gluon Plasma (QGP) is a hot topic in relativistic heavy-ion collisions~\cite{Stephanov,Adams,Asakawa,Koch}. It is suggested that at vanishing or low baryon chemical potential $\mu_B$ and high temperature $T$, the transition is a smooth crossover~\cite{Aoki}. Predictions from effective field theory indicate that at low $T$ and high $\mu_B$, the phase transition may be first-order~\cite{Masayuki,Bowman}, with the CP representing the end point of the first-order transition line. Physicists are currently dedicated to investigating the boundaries of the QCD phase diagram and determining the location of the CP, as these subjects hold considerable scientific interest~\cite{Fodor,Gavai}.

The QCD phase diagram is not yet well understood, either experimentally or theoretically. Lattice QCD calculations are currently limited to scenarios with zero or small baryon chemical potentials due to the sign problem~\cite{latticesign1,latticesign2}. As a result, various phenomenological models, including spin models, have been proposed to study the universality aspect of critical phenomena~\cite{spinmodel1,spinmodel2,spinmodel3,spinmodel4,spinmodel5}. Phase transitions arise from spontaneous symmetry breaking, and systems sharing the same symmetry fall into the same universality class, exhibiting identical critical exponents and potentially similar critical behaviors. It is argued that the QCD critical point belongs to the Z(2) universality class, the same as the 3D Ising model~\cite{Stephanov,Introduction2-1,Introduction2-3,Introduction2-4,Introduction2-5}. By mapping the parametric equation-of-state of the Ising model, one can establish a connection between the phase diagram of the 3D Ising model in the ($T$, $H$) plane and that of QCD in the ($T$, $\mu_{B}$) plane~\cite{Stephanov-map,Parotto}.

Over the past few decades, there has been increasing interest in exploration of geometrical characteristics of phase transitions~\cite{GPhase1,GPhase2,GPhase3}. Geometry holds a unique significance in its ability to provide insight into the underlying mechanisms and behaviors of critical phenomena. By analyzing geometric properties, one can predict critical exponents, scaling laws, and universal properties of a critical system~\cite{GPhase3,Introduction5-1,Introduction5-2,KBinder,FK1}. In particular, percolation theory serves as an optimal framework for this purpose. As a statistical mechanism, percolation theory is dedicated to investigate the behavior of interconnected clusters within various media and how this connectivity affects the macro-scale properties of the system. This theory has demonstrated its utility in detecting the distinctive features of the QCD phase transition. It has been used to investigate the initial phase of deconfinement transition in heavy-ion collisions~\cite{Introduction1-4,Introduction1-5,Introduction1-6}. The method has also been extended to explore the confinement-deconfinement transition of both SU(2)~\cite{SU21,SU22,SU23} and SU(3)~\cite{SU31,SU32} lattice gauge theories. Through a suitable definition of clusters, it is found that the deconfinement transition of gauge theory can be characterized by percolation of clusters within the dominant sector. Additionally, the QGP to hadron crossover transition has been characterized via the temperature dependence of the shear viscosity over entropy density ratio, utilizing the percolation framework within the Color String Percolation Model~\cite{CSPM}.

The studies on phase transitions in the Ising model in the market have commonly employed magnetization to explore the thermal magnetization transition under alterations in temperature or external field. This thermal magnetization transition is characterized by a pivotal Curie temperature ($T_c$). When surpassing $T_c$, the system assumes a disordered phase, but when dropping below it, the system transitions into an ordered phase characterized by a non-zero spontaneous magnetization. The geometrical percolation transition, on the other hand, is related to connectivity of spin clusters~\cite{Introduction1-3}. Envision a lattice of spins where each spin interacts with its nearest neighbors. At a specific critical percolation threshold, denoted as $T_p$, the emergence of percolation cluster, in which the microscopic elements become connected and form a sample-spanning path across the system, is suggested to be an indicator of the occurrence of a continuous percolation phase transition~\cite{Introduction1-7}. 

The investigations on percolation transitions within the Ising model originally concentrate on the Geometrical Spin (GS) clusters, which comprise nearest neighbor spins with the same sign in the lattice. The results of the 2D Ising model show that percolation transition occurs exactly at the critical temperature $T_c$ of the thermal magnetization transition~\cite{Perco2D}. However, a noticeable disparity between these two transitions is observed in three dimensions~\cite{Perco3D}. Consequently, the introduction of Fortuin–Kasteleyn (FK) clusters is proposed~\cite{FKcluster-1,FKcluster-2}, wherein nearest neighboring spins of the same sign are considered to belong to the same cluster with a certain probability. By introducing a parameter of the bond probability, the critical point and the critical exponents of the FK clusters between the thermal and percolation transitions coincide. Nevertheless, precise solutions of the percolation temperature and associate theoretical values of critical exponents concerning the percolation phase transition of the GS clusters within the 3D Ising model remain as open issues.

In traditional statistical physics, the order parameter is commonly used to classify different states of matter and identify phase transitions within a system. However, the intricate nature of robust interactions introduces considerable challenges in ascertaining the order parameter associated with the QCD phase transition and subsequently measuring it through experiments. Recent advancements have proposed integrating machine learning (ML)\cite{ML1, ML2} as a promising approach to explore this complex physical issue\cite{MLReview1, MLReview2}. The advantage of ML methods lies in their strong adaptability and generalization capabilities. They can automatically learn features, handle large-scale data, recognize complex patterns and relationships, and support multi-task learning. This makes ML a powerful tool not only for enhancing efficiency and discovery but also for solving complex problems and achieving intelligent decision-making. Empirical evidence now substantiates that ML techniques can effectively discern the order parameter~\cite{MLIsing5, MLIsing6, MLIsing7, CNNIsing2, CNNIsing4, Feature2DIsing}, learn percolation phase transition in various models~\cite{Per1,Per2} and identify thermal magnetization phase transitions~\cite{spinmodel1,CNNIsing1, MLIsing1, MLIsing2, MLIsing3, MLIsing4} in the Ising model using various ML methodologies. Furthermore, recent study shows that the same neural network can be utilized to identify different phase transitions belonging to the same universality class~\cite{MlUniversality}. This finding enhances the feasibility of using ML methods to investigate phase transitions across a wide range of scenarios.

In this work, we aim to investigate the percolation transition of GS clusters in the 3D Ising model using state-of-the-art ML techniques. Unlike conventional approaches, these methods do not require prior knowledge of order parameters or any additional information. We demonstrate the efficacy of both the unsupervised and semi-supervised ML methodologies in categorizing different phases of the percolation transition. Furthermore, we try to extract the temperature of the percolation threshold and a critical exponent associated with correlation length directly from the output layer of the neural networks. The subsequent sections of this paper are structured as follows: In Section~\ref{PTinModels}, we offer a brief introduction to the percolation transition in the 3D Ising model. Section~\ref{MLmethod} gives an overview of the network algorithms employed in this study. Moving on to Section~\ref{ResultsPCA} and~\ref{ResultsDANN}, we present and discuss the results pertaining to the identification of percolation phase transitions within the Ising model, utilizing the Principal Component Analysis and the Domain Adversarial Neural Network methods, respectively. Finally, in Section~\ref{sec:conclusions}, we summarize our discoveries and provide an outlook for future directions.

\section{Percolation Transition in the 3D Ising Model}\label{PTinModels}
%%%%%%%%%%%%%%%%the defination of Ising %%%%%%%%%%%%%%%%%%%%%%%%%%%%%%%%%%%
Ising model~\cite{Modles-Ising1} is a classic spin model utilized for investigating phase transitions. We consider a three-dimensional cubic lattice composed of $N=L\times L\times L$ sites with periodic boundary conditions. Each site is occupied by a spin, $s_i$. We assume that the spin of individual site can have one of the two states, either spin-up, $s_i = +1$, or spin-down, $s_i = -1$. In this analysis, we only study the 3D Ising model in zero external field, as described by the Hamiltonian 
 \begin{equation}
\mathcal{H}=-J\sum_{\langle ij\rangle}s_{i}s_{j},
 \label{Eq:Energy}
\end{equation}
%\noindent 
where $J$ is known as the coupling constant between two spins $s_i$ and $s_j$. Here we set $J=1$ as the energy unit. The sum is taken only over nearest-neighbor pairs in the grid, and it describes the interaction of the spins with each other. 

%%%%%%%%%%%%%%%%% The introduction of Wolff%%%%%%%%%%%%%%%%%%%%%%%%
The Wolff algorithm, a Monte-Carlo method, is frequently employed to generate equilibrium configurations of the Ising model under varying conditions, such as changes in system size or temperature. In the present investigation, we explore a temperature range spanning from $T=0.02$ to $T=7.92$ with external magnetic field $H=0$. We generate a set of 2000 independent spin configurations for each selected temperature with a given system size.
\begin{figure}
\hspace{-0.3cm}
\includegraphics[scale=0.28]{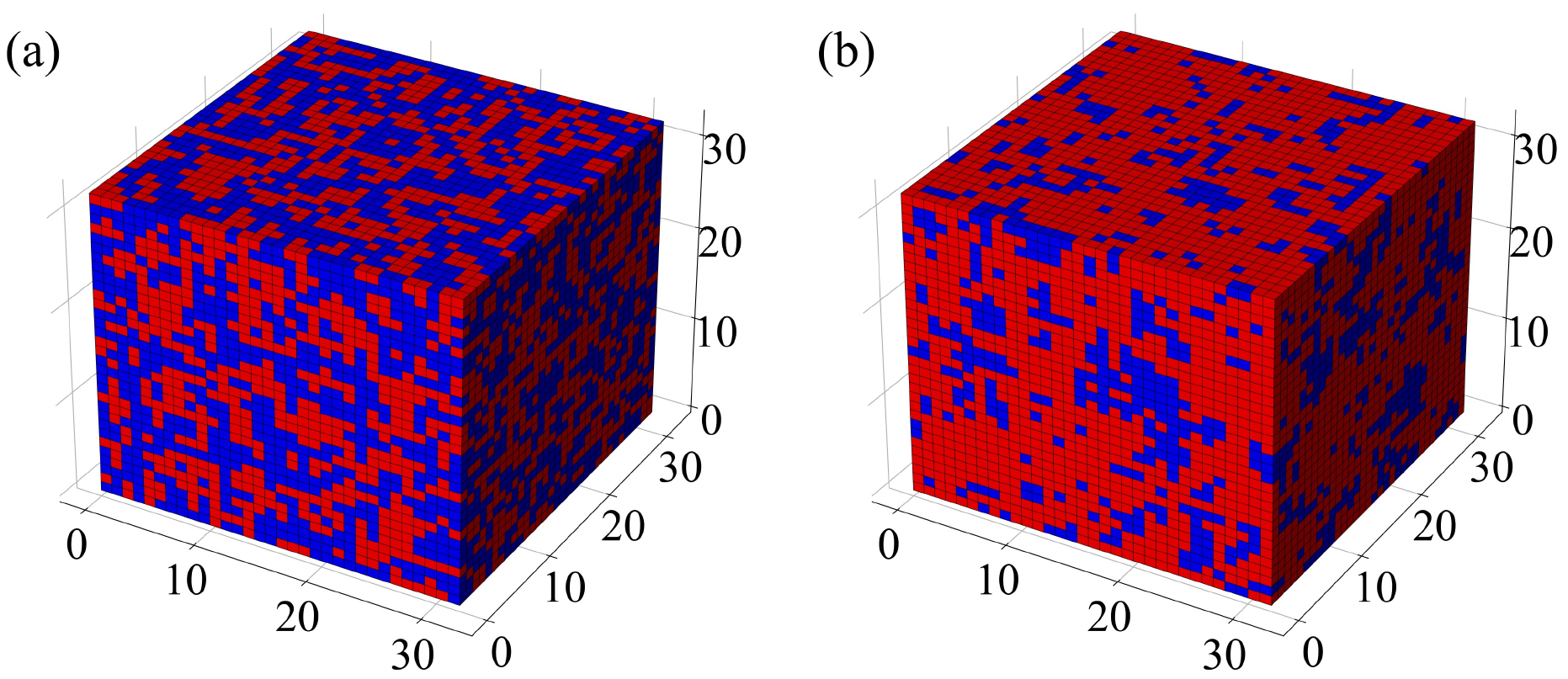}
\caption{(color online) The spin configurations of the 3D Ising model with a lattice size of $L = 32$ are depicted as (a) in the absence of percolation clusters at $T = 7.52$, and (b) in the presence of a percolation cluster at $T = 4.42$.}
\label{cluster:whole}
\end{figure} 

\begin{figure*}
    \centering
    \includegraphics[scale=0.45]{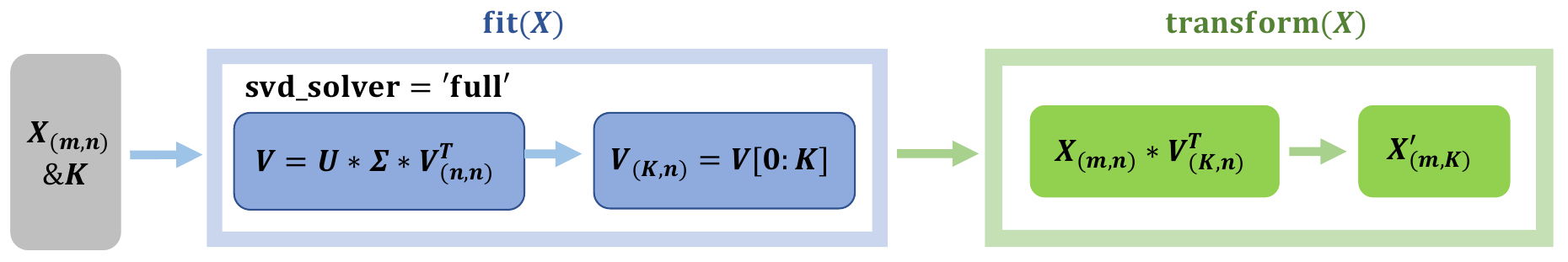}
     \caption{(color online) The PCA architecture used in our analysis.}
    \label{Fig:PCAStructure}
\end{figure*}

 \begin{figure*}
     \centering
     \includegraphics[scale=0.45]{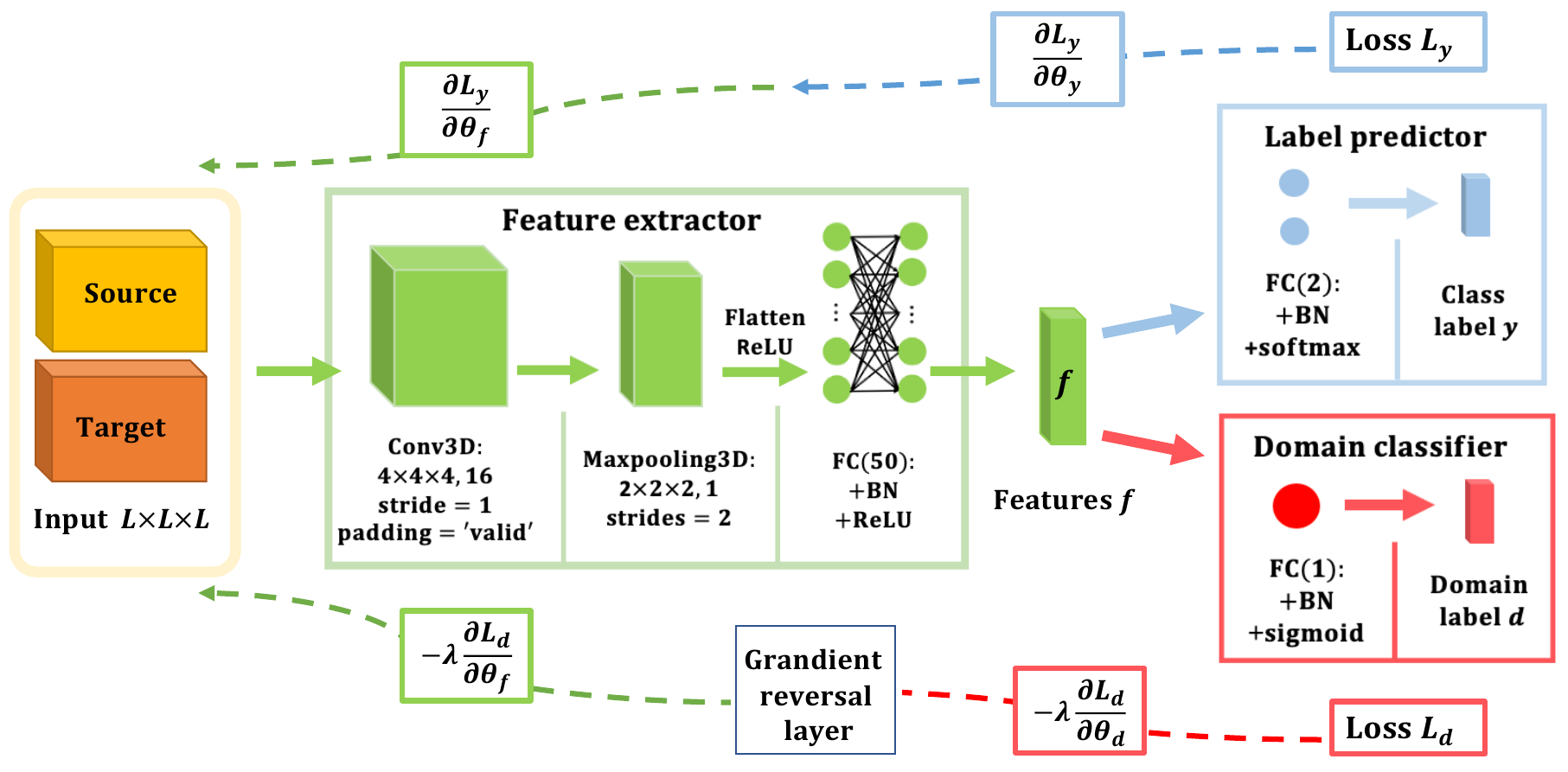}
      \caption{(color online) The DANN architecture used in the analysis.}
     \label{Fig:DANNStructure}
\end{figure*}

In the Ising model, the GS clusters are identified as groups of nearest-neighboring sites with the same spin direction. As the temperature of the system changes, the spins tend to align with their neighbors due to interactions between them. The percolation phase transition infers a significant change in the way clusters of aligned spins form and extend within the whole lattice at a certain critical threshold $T_p$. Beyond the percolation temperature, small clusters of aligned spins are isolated and do not span the entire lattice. However, as the temperature crosses $T_p$, these smaller clusters start to coalesce and connect, leading to the emergence of a percolating cluster that wraps around the whole lattice. This percolating cluster signifies a sudden change in the system behavior, as the alignment of spins becomes correlated over long distances. 

To facilitate an intuitive comprehension, Fig.~\ref{cluster:whole} (a) shows the spin configurations within the 3D Ising model for a system size of $32\times 32\times 32$ at a temperature of $T=7.52$. In this depiction, red lattices represent spin-up orientations, while blue ones denote spin-down. At high temperature, owing to the stochastic distribution of the spin states, no percolation clusters are discernible. Moving to Fig.~\ref{cluster:whole} (b), which illustrates the configuration at a low temperature of $T=4.42$, we observe a predominant cluster colored in red. This cluster spans the whole lattice, signifying the presence of percolation clusters. The percolation transition is of great interest because it often leads to emergent behaviors and critical phenomena, where small changes in a model parameter can lead to drastic alterations in the overall behavior of the system.

\section{ Machine Learning Methodology}\label{MLmethod}
Supervised learning, unsupervised learning, and semi-supervised learning are fundamental paradigms in machine learning~\cite{Models-ML}, each with different approaches and applications. Supervised learning involves training a model on labeled data, where input-output pairs are provided, enabling the model to learn patterns and make accurate predictions on new, unseen data. Unsupervised learning, on the other hand, deals with unlabeled data, aiming to uncover inherent structures, clusters, or relationships within the data. Semi-supervised learning merges elements of both, incorporating labeled and unlabeled data to enhance model performance. By leveraging the small amount of labeled data alongside the larger pool of unlabeled data, semi-supervised learning strikes a balance between efficiency and accuracy, making it valuable when acquiring fully labeled datasets is expensive or time-consuming. 

To enhance the versatility and future applicability of our approach, we use both the unsupervised Principal Component Analysis (PCA) method and the semi-supervised Domain Adversarial Neural Network (DANN) methodology in this study. It allows to extend the possible utility of our methods to the analysis of experimental data, ensuring a broader scope and improved adaptability for future applications in high energy physics.

PCA is among the most widely utilized multivariate techniques~\cite{Models-PCA1}, with its origins tracing back to Pearson's pioneering work in statistics~\cite{Models-PCA2}. Pearson's formulation involves identifying optimal lines and planes to closely align with point distributions in space. PCA aims to simplify complex datasets by identifying a new set of orthogonal axes, called principal components, that capture the most significant variations in the data. These components are ordered in terms of the amount of variance they explain, allowing for the reduction of high-dimensional data into a lower-dimensional space while retaining as much relevant information as possible. Functioning as a classic example of unsupervised learning, PCA finds extensive application in tasks such as data clustering and dimensionality reduction~\cite{MLIsing5,MLIsing6,Models-PCA5}.

The sketch in Fig.~\ref{Fig:PCAStructure} illustrates the network architecture employed by the PCA algorithm in this investigation. It can be segmented into two main components: the fit stage and the transform stage. To begin with, the network takes as input the spin configurations of the data denoted as $X_{(m,n)}$, along with the predetermined number of principal component features, denoted as $K$. In the fit stage, PCA computes the mean and covariance matrix for the dataset $X_{(m,n)}$. Subsequently, the Singular Value Decomposition (SVD) decomposes the covariance matrix into eigenvalues and their corresponding eigenvectors, which are denoted as $V$. Fig.~\ref{Fig:PCAStructure} presents the mathematical expression of SVD as $U \ast \Sigma \ast V{}^{T}_{\left( n,n\right)}$. In this expression, $\Sigma$ contains the singular values of the original matrix, while $U$ and $V_{\left( n,n\right)}$ represent the left and right singular vectors of the original matrix, respectively. $V{}^{T}_{\left( n,n\right)}$ denotes the transpose of $V_{\left( n,n\right)}$. Subsequently, based on the predefined number of features $K$, we retain the first $K$ rows of the feature matrix $V$, represented as $V_{(K,n)} = V[0:K]$. Shifting to the transform stage, the network executes a projection of the original data $X_{(m,n)}$ onto the selected principal components $V{}^{T}_{\left( K,n\right)}$. As a result, the data is transformed into a reduced-dimensional representation, denoted as $X^{\prime}_{m,K}$. This accomplishment effectively realizes the objectives of both dimensionality reduction and feature extraction.

DANN~\cite{Modles-DANN1} is a specific type of transfer learning~\cite{Models-DANN2,Models-DANN3} that emphasizes the mapping relationship between labeled source domain data and unlabeled target domain data, rather than the clustering and dimensionality reduction features provided by PCA. DANN uniquely integrates deep feature learning and domain adaptation, enabling the classification decision to be based on both discriminative and domain-invariant features for accurate data classification. By leveraging DANN, the classification process can effectively utilize features that are both informative and immune to domain variations.

The overall structure of DANN is shown in Fig.~\ref{Fig:DANNStructure}, comprising three main components: a feature extractor (green), a label predictor (blue), and a domain classifier (red). The feature extractor captures informative features from the input data, transforming them into the feature vector $f$. It is composed of a convolutional layer, a max-pooling layer and a fully connected hidden layer. In the convolutional layer, there are 16 filters, each measuring $4 \times 4 \times 4$ with a stride of 1, applied to the input data to generate feature maps. The max-pooling layer uses a filter of size $2 \times 2 \times 2$ with strides of 2 to reduce the dimensionality of the feature maps. These feature maps are then flattened and passed to a fully connected (FC) layer containing 50 neurons. Additionally, Batch Normalization (BN) and ReLU activation are applied to prevent overfitting and speed up the training process. Once the feature extractor has extracted feature vectors $f$ from both the source and target domain data, the network forwards these vectors $f$ to the label predictor and the domain classifier. The label predictor consists of a fully connected layer with 2 neurons, applying BN and softmax activation. Its output is expressed as a vector ($P_0$, $P_1$), indicating the probabilities of the input data belonging to different event types. The domain classifier comprises a fully connected layer with 1 neuron, utilizing BN and sigmoid activation. It outputs the vector $P_d$, determining whether the feature vector $f$ originates from the source domain or the target domain; if it is from the source domain, $P_d$ = 0; if from the target domain, $P_d$ = 1. In our neural network architecture, the feature extractor and label predictor form a conventional feedforward neural network. Additionally, the feature extractor constitutes an adversarial network by connecting a gradient reversal layer to the domain classifier. The loss function for the domain adversarial network comprises two components: one for the label predictor and the other for the domain classifier. The loss function of the network is defined as: $L(\theta_f, \theta_{y}, \theta_{d})= L_{y}( \theta_{f} , \theta_{y} )-L_{d}( \theta_{f} , \theta_{d} ),$ where $y$, $f$, and $d$ refer to the labels of the three components of DANN, and $\theta$ represents the internal parameters of the network. The learning rate $\lambda$ is set to 0.0005. During the iterative training process, the $Adam$ optimizer in TensorFlow 2.4.1 is utilized to minimize the loss functions for each component of the network. Once training is complete, the trained model can be employed to predict unlabeled samples and generate prediction results. This approach utilizes both labeled and unlabeled samples to train different parts of the model, facilitating comprehensive training and optimization of the model.
%%%%%%%%%%%%%%%%%%%%%%%%%%%%%%%%%%%%%%%%%%%%%%%%%%%%%%%%%%%%%%%%%
\section{ML the percolation transitions in the Ising model by PCA}\label{ResultsPCA}
%%%%%%%%%%%%%%%%%%%%%%%%%%%%%%%%%%%%%%%%%%%%%%%%%%%%%%%%%%%%%%%%%

The geometrical percolation transition typically involves several key quantities, such as percolation strength, the largest cluster size, average cluster size, as discussed in reference~\cite{typicalvar}. The cluster size, denoted as $S$, is determined by the count of spin sites it encompasses. To ensure sufficient data for subsequent calculations, our primary focus is on the largest cluster size, referred to as $S_{max}$, which is influenced by both temperature and volume of the system.

\begin{figure}
\hspace{-0.8cm}
\includegraphics[scale=0.4]{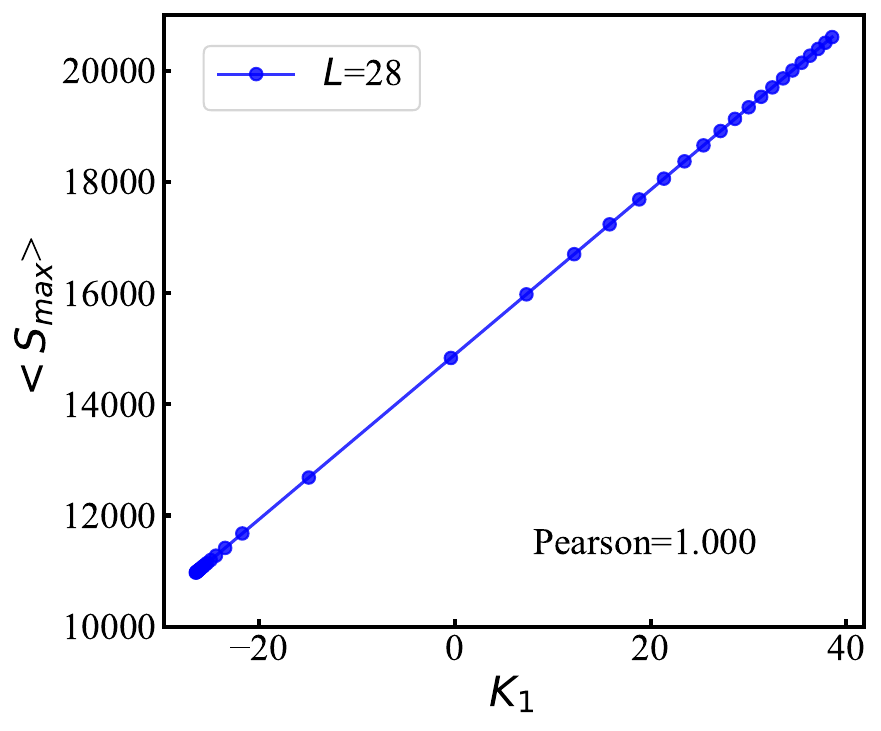}
\caption{(color online) The average size of the largest clusters as a function of the output of the first principal component in PCA with $L = 28$.}
\label{Fig:Pearson}
\end{figure}

It has been verified that $S_{max}$ plays a crucial role in understanding the behavior of systems undergoing a phase transition from a disconnected state to a connected state~\cite{spinmodel3}. The appearance of a percolating cluster serves as a critical indicator of a phase transition within the system. Along with other extensive variables such as susceptibility and correlation length, the size of the largest cluster obeys scaling laws near the critical threshold. Understanding the size and structure of the largest cluster is essential for gaining insights into the critical behavior of percolating systems and characterizing overall connectivity properties.

In the thermal magnetization phase transition of the Ising model, it has been found that the first principal component ($K_1$), defined as the direction that maximizes variance in the data,  captures the highest level of variability in the dataset~\cite{CNNIsing4}. The results derived from $K_1$ confirm its linear relationship with magnetization within a finite system size. This implies that the first principal component of the input data can effectively capture and learn about magnetization, which is the characteristic order parameter for the thermal magnetization transition in the 3D Ising model. To assess the capacity of the unsupervised ML method for learning and identifying important features associated with the largest cluster size from the input data in the percolation transition, we choose PCA methodology with one principal component. 

For the implementation of PCA network, we employ the PCA class methods from the scikit-learn library 1.0.1 in Python 3.7.11, focusing on data processing with the relevant functions. The key hyperparameters of our PCA network include the dimensionality reduction parameter $K_1$ and the singular value decomposition method set as $svd\_solver = full$. Upon configuring these hyperparameters, we employ the $pca.fit()$ function to conduct principal component analysis on the input data. This function fits the PCA model to the data, enabling the computation of principal components and explained variance ratios. It involves learning the transformation parameters from the data necessary for dimensionality reduction. Subsequently, the $pca.transform()$ function is used to reduce the dimensionality of data by projecting it onto the new feature space defined by the principal components. This process effectively reduces the dimensionality of data while preserving the essential information captured by the principal components.

We generate event samples for a given system size of $L = 28$ and cover a temperature range spanning from $T=3.52$ to $T=5.52$. The spin configurations of the largest cluster, which means we retain only the spins in that cluster and set all other spins to 0, are utilized as input data for each sample of the 3D Ising model for the PCA network architecture as depicted in Fig.~\ref{Fig:PCAStructure}. Subsequently, we compute the average sizes of the 2000 largest clusters at each temperature point and conduct a Pearson correlation analysis with respect to the first principal component obtained from PCA. The correlation between the average size of the largest clusters $\langle S_{max}\rangle$ and the first principal component $K_1$ of the PCA is graphically illustrated in Fig.~\ref{Fig:Pearson}. Our findings reveal a linear relationship between $\langle S_{max}\rangle$ and $K_1$. The computed Pearson correlation coefficient of 1.0 suggests a strong positive correlation between $\langle S_{max}\rangle$ and $K_1$.

\begin{figure}
\hspace{-0.53cm}
\includegraphics[scale=0.335]{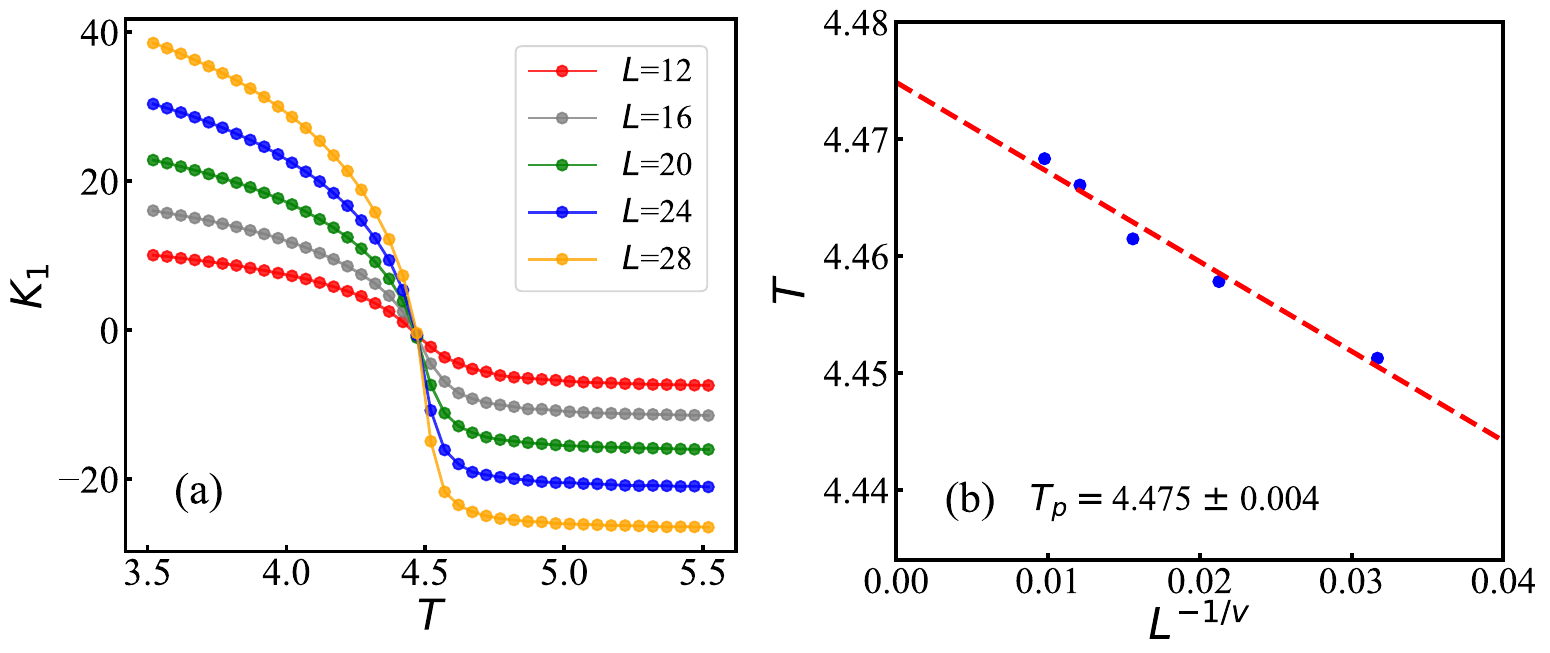}
\caption{(color online)(a) The first principal component of PCA as a function of temperature for five different system sizes. (b) Finite-size scaling analysis to determine percolation critical threshold based on the first principal component.}
\label{Fig:K1 for PCA}
\end{figure}

 \begin{table*}
     \centering
     \includegraphics[scale=0.45]{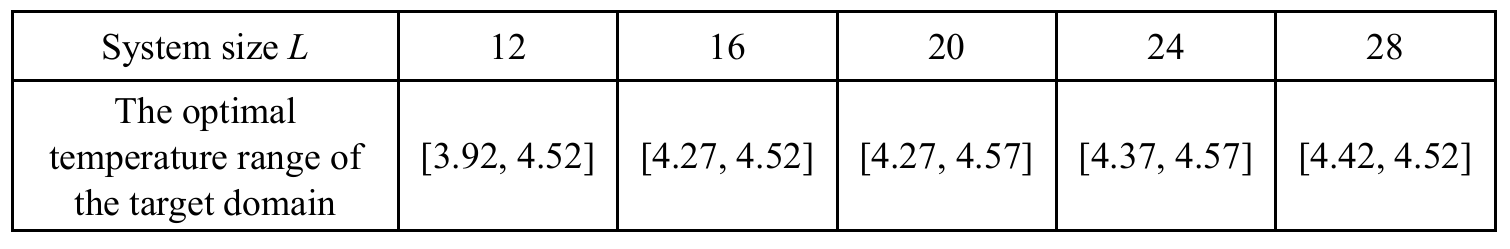}
      \caption{Selection of optimal temperature range of target domain in DANN.}
     \label{Fig:DANNselect}
\end{table*}

As we know, the first principal component represents the newly defined coordinate axis within PCA. This axis encapsulates the highest amount of information and exhibits the greatest power in distinguishing patterns in the data. It plays the most significant role in accounting for variations in the data and excels at elucidating data changes. The observed strong positive correlation between the average size of the largest clusters $S_{max}$ and $K_1$ provides further confirmation that the largest cluster carries a substantial amount of critical system information, and this information can be effectively acquired from the first principal component of PCA in the geometric percolation phase transition.

We will now explore the capability of PCA in identifying different phases and investigating the critical threshold associated with the percolation transition within the 3D Ising model. As detailed in Sec.~\ref{PTinModels}, the Ising model undergoes a geometric percolation phase transition at a critical threshold $T_p$. This transition separates a disconnected state at high temperatures from a fully connected state at low temperatures. To achieve this, we establish a PCA network to conduct unsupervised learning directly on samples of spin configurations of the largest clusters. 

The numerical results obtained from five different system sizes and spanning the temperature range $T\in \left[ 3.52,5.52\right]$ are illustrated in Fig.~\ref{Fig:K1 for PCA} (a). The first principal component $K_1$ initially exhibits a gradual decrease as the temperature increases for all sizes considered. This is followed by a sharp drop as the temperature approaches a specific point, ultimately reaching equilibrium at higher temperatures. Notably, the $K_1$ values for different sizes intersect at a particular temperature, corresponding to the critical threshold of the percolation phase transition. This demonstrates the effectiveness of PCA as a methodology for classifying the two different phases of the percolation transition in the 3D Ising model.

We conduct five independent iterations of the PCA network, each with a different size of $L =$ 12, 16, 20, 24, 28. The pseudo-critical threshold $T$ for each size  corresponds to the point where $K_{1} =0 $. Subsequently, we employ a finite-size scaling analysis to estimate the percolation transition temperature in the infinite $L$ limit, using information derived from the first principal component~\cite{CNNIsing4,3DIsingCriT3}. Followed by the same methods as used in Refs.~\cite{expofit1,expofit2}, the percolation temperature $T_p$ is achieved by extrapolating through the fit of $|T - T_p| \sim L^{-1/\nu}$ as the limit $1/L$ approached zero, as illustrated in Fig.~\ref{Fig:K1 for PCA} (b). The statistical errors are estimated using the standard deviation and are found to be smaller than the size of the data points. The percolation temperature determined from the fitting process yields $T_p = 4.475\pm 0.004$. We would point out that this critical threshold for the percolation transition in the 3D Ising model is slightly lower than the critical temperature for the thermal magnetization phase transition, which is $T_c = 4.5115\pm 0.0001$ as reported in references~\cite{NewTc1,NewTc2}. This finding is qualitatively consistent with the results in reference~\cite{Perco3D, Results-1}.

%%%%%%%%%%%%%%%%%%%%%%%%%%%%%%%%%%%%%%%%%%%%%
\section{ML the percolation transitions by DANN}\label{ResultsDANN}
 
Domain Adversarial Neural Network is a deep learning technique mainly applied in the domain adaptation field. The primary advantage of DANN is its ability to adapt a machine learning model from one domain to another, particularly when the source and target domains have different data distributions. This neural network is trained in such a way that the feature representations of the two domains become indistinguishable to the domain classifier. 

In our analysis, we label percolation phenomenon at extremely low temperatures as phase $'1'$ and those with no percolation phenomenon at exceedingly high temperatures as phase $'0'$ in the Ising model. Consequently, we designate the spin data of the largest clusters at low and high temperatures as the source domain data during DANN network training, with the unlabeled one in the intermediate temperature range regarded as the target domain data. Detailed network architectures and the training process of DANN are illustrated in Fig.~\ref{Fig:DANNStructure}. To establish the optimal temperature range for each scale, we employ a technique involving the fixation of either low-temperature or high-temperature labels while decreasing the high-temperature labels or increasing the low-temperature labels, respectively. The specific selection of the temperature range for target domain data at various system sizes is outlined in Table~\ref{Fig:DANNselect}. The source domain data can then be annotated with suitable labels according to the target domain range. In this process, DANN learns the mapping relationship between the source domain data and the target domain data. The ultimate label predictor proves effective at accurately forecasting the percolation classification of unlabeled data in the target domain through the utilization of domain adaptation and back-propagation techniques. 

\begin{figure*}
    \centering
    \includegraphics[scale=0.54]{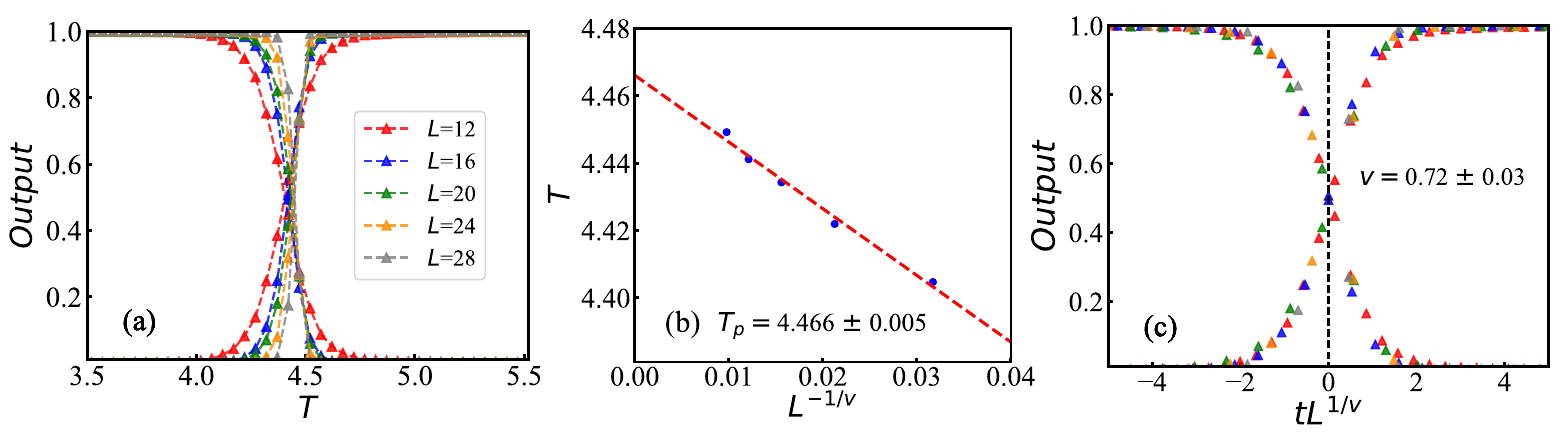}
     \caption{(color online) (a) The output layer averaged over test sets of DANN as a function of temperature for five different system sizes. (b) Finite-size scaling analysis to determine percolation critical threshold. (c) Data collapse of the average output layer as a function of $tL^{1/v}$, where $t = (T-T_{p})/T_{p}$ is the reduced temperature.}
    \label{Fig:Out of DANN}
\end{figure*}

After training the DANN on the optimal domain, we assess samples at various temperatures, and then use DANN to predict the probability of each configuration belonging to phase $'1'$ or phase $'0'$. The numerical results obtained at various system sizes at a vanishing magnetic field are illustrated in Fig.~\ref{Fig:Out of DANN} (a). The average outputs of different sizes cross at a specific temperature corresponding to the percolation temperature. It infers that the DANN can successfully classify the two different phases in the percolation transition in the Ising model. 

To determine the percolation critical threshold, we conduct DANN training at five different sizes and identify the pseudo-critical threshold $T$ as the intersection point of  the two curves corresponding to phase $'0'$ and phase $'1'$ for each size. The percolation threshold $T_p$ is achieved by extrapolating through the fit of $|T - T_p| \sim L^{-1/\nu}$ as the limit $1/L$ approached zero, as illustrated in Fig.~\ref{Fig:Out of DANN} (b). The obtained critical threshold for the percolation transition, $T_p = 4.466\pm 0.005$, is in agreement with the result from the PCA network within errors. This temperature is slightly lower than the critical temperature associated with the thermal magnetization phase transition in the 3D Ising model.

The critical exponents associated with the 3D Ising model play a crucial role in characterizing the behavior of various thermodynamic and correlation functions as the system approaches the critical point. They are considered universal in nature, $i. e.$, they remain unchanged regardless of the specific characteristics of the physical system. Among these exponents, the correlation length exponent $\nu$ holds significance as it reveals how the correlation length undergoes a pronounced divergence near the critical point. This divergence is described by the relationship between the correlation length $\xi$ and temperature, expressed as $\xi\sim|T - T_p|^{-\nu}$. In the case of a finite system, it is expected that the correlation length scales proportionally to the size of the system. Consequently, one can establish a connection between the temperature and the system size, characterized by the expression $|T - T_p| \sim L^{-1/\nu}$. 

To extract the critical exponent $\nu$, we employ the technique of data collapse~\cite{PRE1,PRE2} and the results are shown in Fig.~\ref{Fig:Out of DANN} (c). What becomes evident from the figure is that as different system sizes are considered, they exhibit a compelling convergence with  $\nu = 0.72 \pm 0.03$. This value quantitatively agrees with the previously measured cluster size heterogeneity result of $\nu \approx 0.75$ for geometric clusters in the 3D Ising model~\cite{3DIsingnu}. Earlier studies on FK clusters~\cite{FKCluster1,FKCluster2} demonstrated that the calculated correlation length exponent $\nu$ is consistent with that of the thermal magnetization phase transition, reported as $\nu = 0.629 912(86)$ using the MC method~\cite{NewTc1}. Our findings indicate that the extracted correlation length exponent for GS clusters surpasses that of the thermal magnetization phase transition in the 3D Ising model.

%%%%%%%%%%%%%%%%%%%%%%%%%%%%%%%%%%%%%%%%%%%%%%%%%%%%%%%%%%%%%%%%%%%%%%%%%%%%%%%
\section{Conclusions and Outlook}\label{sec:conclusions}
%%%%%%%%%%%%%%%%%%%%%%%%%%%%%%%%%%%%%%%%%%%%%%%%%%%%%%%%%%%%%%%%%%%%%%%%%%%%%%%
In summary, we employ both unsupervised and semi-supervised machine learning techniques to investigate percolation phase transitions of the geometrical spin clusters within the 3D Ising model. We find a linear correlation between the average size of the largest clusters and the primary component of PCA, suggesting that the largest clusters contain sufficient information pertaining to the percolation transition. We use two distinct machine learning approaches by utilizing the spin configurations from the largest clusters as input data, and apply the finite-size scaling method to estimate the critical threshold of the percolation phase transition. Both of these machine learning methodologies effectively classify disconnected states at higher temperatures and fully connected states at lower temperatures. As a result, we determine the critical threshold for the percolation transition to be $T_p = 4.475\pm 0.004$ and $T_p = 4.466\pm 0.005$ by using PCA and DANN, respectively. These temperatures are found to be a little bit lower than the critical temperature for the thermal magnetization phase transition. The extracted correlation length exponent is found to be $\nu = 0.72 \pm 0.03$, which is consistent with the cluster size heterogeneity result for geometric clusters. This value is  greater than the critical exponent  associated with the thermal magnetization phase transition in the 3D Ising model.

Percolation theory has become a powerful tool for investigating phase transitions in various physical systems. The machine learning methods developed in this study have the remarkable ability to identify phase transitions and extract critical temperature and critical exponents with minimal or even no training data. This capability allows for the convenient application in the investigation of criticality in high energy experimental data or other Monte Carlo models, which could help to explore the underlying physical mechanisms governing the QCD phase transition in heavy-ion collisions.

%%%%%%%%%%%%%%%%%%%%%%%%%%%%%%%%%%%%%%%%%%%%%%%%%%%%%%%%%%%%%%%%%%%%%%%%%%%%%%%
\section*{Acknowledgments}
%%%%%%%%%%%%%%%%%%%%%%%%%%%%%%%%%%%%%%%%%%%%%%%%%%%%%%%%%%%%%%%%%%%%%%%%%%%%%%%
We are grateful to Prof. Mingmei Xu, Lizhu Chen and Dr. Feiyi Liu for fruitful discussions and comments. We further thank Prof. Hengtong Ding for providing us with computing resources. The numerical calculation have been performed on the GPU cluster in the Nuclear Science Computing Center at Central China Normal University (NSC3). This work is supported by the National Key Research and Development Program of China (No. 2024YFE0110103 and No. 2022YFA1604900), the National Natural Science Foundation of China (No. 12275102) and the Chinese Scholarship Council Grant No. 202306770049. 

R.G. and X.L. contributed equally to this work.

%\bibliography{ref}% BiTex form
%\include{ref}

%%%%%%%%%%%%%%%%%% Citation %%%%%%%%%%%%%%%%%%%%%%%%%

\end{document}